\documentclass[structabstract]{raa_rb}
\usepackage{soul}
\usepackage{color}
\usepackage{booktabs}
\usepackage{graphicx,times}
\usepackage{natbib,amssymb}
\usepackage{epstopdf,epsfig}
\usepackage{float}
\usepackage{amsmath}
\usepackage{ulem}
\usepackage{amsmath}

\usepackage{mathtools}

\providecommand{\bm}[1]{\mbox{\boldmath$#1$\unboldmath}}
\def\romeI{$\mathrm{I}$}

\begin{document}
   \title{The progenitors of Type Ia supernovae with asymptotic giant branch donors}

   \volnopage{ {\bf 2023} Vol.\ {\bf X} No. {\bf XX}, 000--000}
   \setcounter{page}{1}

   \author{Lu-Han Li \inst{1,2,3,4},
	       Dong-Dong Liu \inst{1,2,3}
          \and
               Bo Wang \inst{1,2,3}
          }

   \institute{Yunnan Observatories, Chinese Academy of Sciences, Kunming 650216, China;
          {\it liudongdong@ynao.ac.cn; wangbo@ynao.ac.cn}\\          
          \and
         Key Laboratory for the Structure and Evolution of Celestial Objects, Chinese Academy of Sciences, Kunming 650216, China
         \and
         International Centre of Supernovae, Yunnan Key Laboratory, Kunming 650216, P. R. China\\    
         \and
         University of Chinese Academy of Sciences, Beijing 100049, China\\
              }

   \date{Received ; accepted}

\abstract {Type \romeI a supernovae (SNe \romeI a) are among the most energetic events in the Universe. 
They are excellent cosmological distance indicators due to the remarkable homogeneity of their light curves.
However, the nature of the progenitors of SNe \romeI a is still not well understood. 
In the single-degenerate model, 
a carbon-oxygen white dwarf (CO WD) could grow its mass by accreting material from an asymptotic giant branch (AGB) star, leading to the formation of SNe \romeI a when the mass of the WD approaches to the Chandrasekhar-mass limit, known as the AGB donor channel.
In this channel, previous studies mainly concentrate on the wind-accretion pathway for the mass-increase of the WDs.
In the present work, we employed an integrated mass-transfer prescription for the semidetached WD+AGB systems, and evolved a number of WD$+$AGB systems for the formation of SNe \romeI a through the Roche-lobe overflow process or the wind-accretion process.
We provided the initial and final parameter spaces of WD$+$AGB systems for producing SNe \romeI a.
We also obtained the density distribution of circumstellar matter at the moment when the WD mass reaches the Chandrasekhar-mass limit.  
Moreover, we found that the massive WD$+$AGB sample AT$\,$2019qyl can be covered by the final parameter space for producing SNe \romeI a, indicating that AT$\,$2019qyl is a strong progenitor candidate of SNe \romeI a with AGB donors.
 }

\titlerunning{The progenitors of SNe Ia with AGB donors}

\authorrunning{L.-H. Li et al.}

\maketitle
\section{INTRODUCTION} \label{1. INTRODUCTION}

Type \romeI a supernovae (SNe \romeI a) have strong Si$\mathrm{II}$ absorption lines, but no H and He lines near the maximum luminosity in their spectrum (\citealt{1997ARA&A..35..309F}). 
Due to the homogeneity of the SN \romeI a light curves, they are good distance indicators and used for precise distance measurements in cosmology, revealing the current accelerating expansion of the Universe most possibly driven by dark energy (e.g. \citealt{1998AJ....116.1009R,1999ApJ...517..565P,2011NatCo...2..350H}).  
It has been suggested that the local Hubble constant could be accurately measured if the Hubble flow samples of SNe \romeI a and the calibrations of Cepheid variables could be well combined (\citealt{2019ApJ...876...85R,2022ApJ...934L...7R}).

SNe \romeI a are thought to be the thermonuclear explosion of carbon-oxygen white dwarfs (CO WDs) with masses close to the Chandrasekhar-mass limit ($M_\mathrm{{Ch}}$) in close binaries (e.g. \citealt{1960ApJ...132..565H,1984ApJ...286..644N}). 
However, the nature of progenitor system of SNe \romeI a, especially the donor is still unclear. 
In the past decades, many progenitor models have been proposed, in which the most popular models are the single-degenerate (SD) model and the double-degenerate (DD) model. 
(1) In the SD model, a CO WD increases its mass by accreting H-/He-rich material from a non-degenerate donor, and explodes as an SN \romeI a when its mass approaches to $M_\mathrm{{Ch}}$.
Typically, the non-degenerate donor of the WD can be a main-sequence (MS) star , a sub-giant, a red giant (RG), an asymptotic giant branch (AGB) star, or a He star (e.g.  \citealt{1973ApJ...186.1007W,1982ApJ...253..798N,1997A&A...322L...9L,2000A&A...362.1046L,2004MNRAS.350.1301H,2006MNRAS.368.1095H,2009MNRAS.395..847W,2010RAA....10..235W}). 
(2) In the DD model, a CO WD merges with another CO WD driven by the gravitational wave radiation, which may lead to the formation of an SN \romeI a if their total mass is larger than $M_\mathrm{{Ch}}$ (e.g. \citealt{1984ApJS...54..335I,1984ApJ...277..355W,1998MNRAS.296.1019H,2016MNRAS.461.3653L,2017A&A...606A.136L,2018MNRAS.473.5352L}).
In addition, there are some other progenitor models to explain the observed variety of SNe \romeI a, such as the core-degenerate (CD) model, the hybrid CONe model, the common-envelope wind model, the double WD collision model (for recent reviews, see \citealt{2018RAA....18...49W,2018SCPMA..61d9502S,2018PhR...736....1L}). 

Observationally, there are some massive WD$+$AGB systems that are SN \romeI a progenitor candidates, such as V407\,Cyg, AT\,2019qyl and TUVO-22albb.
(1) V407\,Cyg is considered as a symbiotic star containing a mira donor and a massive WD (e.g.  \citealt{2003MNRAS.344.1233T,2003AstL...29..405T,2012BaltA..21...68H}), which have almost the widest orbit among symbiotic stars with an orbital period of $43\,\mathrm{yr}$ (\citealt{1990MNRAS.242..653M}).
The WD in V407\,Cyg is at least $1.2\, \rm{M_{\odot}}$ (\citealt{2010arXiv1011.5657M}), and  
 may be as massive as $1.35\mathrm{-}1.37\, \rm{M_{\odot}}$ (\citealt{2012BaltA..21...68H}).
(2) AT\,2019qyl is a nova with an O-rich AGB donor in the nearby Sculptor Group galaxy NGC 300 (\citealt{2021ApJ...920..127J}).
\cite{2021ApJ...920..127J} estimated that the allowed range of the AGB star mass in AT\,2019qyl is $M_{\mathrm{2}} = 1.2\mathrm{-}2.0\, \mathrm{M_{\odot}}$, with the best-fitting value is $M_{\mathrm{2}} = 1.2\, \mathrm{M_{\odot}}$ and the orbital period $P \gtrsim 1800\, \mathrm{d}$ by assuming the mass ratio to be 1.
In the present work, we found that the estimated parameters of AT\,2019qyl can be covered based on the WD$+$AGB channel, also known as the AGB donor channel.
(3) TUVO-22albb, located in the nearby spiral galaxy NGC 300, is a probable very fast nova discovered by \cite{2022arXiv221006057M} in their Transient UV Object project, and its donor has been suggested to be an AGB star by further comparison with colour$-$magnitude diagram.

The WD$+$AGB systems will form dense circumstellar medium (CSM) via the mass-loss of the AGB wind or the Roche-lobe overflow (RLOF) process.
The interaction between SN ejecta and pre-explosion CSM can generate electromagnetic radiation in X-ray and radio bands (\citealt{1982ApJ...259..302C}).
Detecting the signal from the interaction between explosive ejecta of the SN and CSM can help us distinguish different progenitor systems.
In the observations, the supernova remnant (SNR) of SN 1604, also known as Kepler's SNR, is located relatively high above the Galactic plane.
SN 1604 is considered as an SN \romeI a because of its prominent Fe-L emission and relatively little oxygen emission (\citealt{1999PASJ...51..239K,2007ApJ...668L.135R}).
\cite{2012A&A...537A.139C} suggested that a WD and a $4\mathrm{-}5\, \mathrm{M_{\odot}}$ AGB donor provided a possible pathway to explain the characteristics of Kepler's SNR through hydrodynamical simulations.

Some previous studies suggested that the AGB donor channel can produce SNe \romeI a through the wind-accretion, but it is relatively difficult to produce SNe \romeI a from stable RLOF process (e.g. \citealt{1997A&A...322L...9L,1998ApJ...497..168Y,2004MNRAS.350.1301H}).
The main reason for this is that previous studies usually assumed that the exceeding mass of the donor should be immediately transferred to the accretor as soon as the donor exceeds its Roche-lobe, which may overestimate the mass-transfer rate when the mass donor is a giant star and thus prevent the WD from increasing its mass to $M_\mathrm{{Ch}}$ (for more discussions see \citealt{2019A&A...622A..35L}).
Recently, \cite{2019A&A...622A..35L} adopted an integrated RLOF mass-transfer prescription described in \cite{2010ApJ...717..724G} to investigate the mass-transfer process of semidetached WD$+$RG systems and provided a significantly enlarged parameter space for producing SNe \romeI a.

In the present work, we adopted the integrated RLOF mass-transfer prescription of \cite{2010ApJ...717..724G} for the mass-transfer process of the semidetached WD$+$AGB systems.
We provided the parameter space of WD$+$AGB systems for the production of SNe \romeI a both through the mass-transfer of RLOF and wind-accretion. 
In Sect.$\,$2, we describe the numerical methods and basic assumptions employed in this work.
The corresponding results are presented in Sect.$\,$3. 
Finally, a discussion and summary are given in Sect.$\,$4.

\section{NUMERICAL METHODS} \label{2. NUMERICAL METHODS}
By using the Eggleton stellar evolution code (\citealt{1973MNRAS.163..279E}), we evolve a large number of WD$+$AGB star systems, in which the WDs are treated as point mass.
We adopt the typical Pop I composition (H fraction $X$ = 0.7, He fraction $Y$ = 0.28, and metallicity $Z$ = 0.02) for the initial MS models.
In this work, we consider the mass-transfer both through RLOF and wind-accretion.
When the mass of WDs grows up to 1.378 $\mathrm{M_{\odot}}$, we assume that WDs would explode as SNe \romeI a.
We consider the angular momentum loss due to the mass-loss, including the stellar wind of the donors and the mass-loss around the WDs through optically thick wind or nova outburst.

\subsection{The Roche-lobe overflow process}

We investigated the mass-transfer rate in semidetached WD$+$AGB systems by the integrated RLOF mass-transfer prescription shown in \cite{2010ApJ...717..724G}, written as

\begin{equation}
\dot{M}_2=-\frac{2 \pi R_{\mathrm{L}}^3}{G M_2} f(q) \int_{\phi_{\mathrm{L}}}^{\phi_{\mathrm{s}}} \Gamma^{1 / 2}\left(\frac{2}{\Gamma+1}\right)^{\frac{\Gamma+1}{2(\Gamma-1)}}(\rho P)^{1 / 2} \mathrm{~d} \phi,
\end{equation}
in which $R_{L}$ is the effective Roche-lobe radius of the donor, $G$ is the gravitational constant, $M_{2}$ is the donor mass, the mass ratio $q = M_{\mathrm{2}}/M_{\mathrm{WD}}$, $\Gamma $ is the adiabatic index, $\rho$ is the local gas density, and $P$ is the local gas pressure. The upper and lower limits of integral are stellar surface potential energy ($\phi_{S}$) and the Roche-lobe potential energy ($\phi_{L}$), respectively. The integration over potential $\phi$ is approximately expressed as follows:
\begin{equation}
\mathrm{d} \phi=G M_2 R^{-2} \mathrm{~d} R,
\end{equation}
where R is the donor radius. The combined coefficient $f(q)$ is a slowly varying function of the mass ratio $q$:
\begin{equation}
f(q) \equiv \frac{q}{r_{\mathrm{L}}^3(1+q)} \frac{1}{\left[a_2\left(a_2-1\right)\right]^{1 / 2}},
\end{equation}
where $a_2$ is defined as
\begin{equation}
a_2=\frac{q}{x_{\mathrm{L}}^3(1+q)}+\frac{1}{(1+q)\left(1-x_{\mathrm{L}}\right)^3},
\end{equation}
in which $x_L$ is accurately approximated as
\begin{equation}
x_{\mathrm{L}}=\left\{\begin{array}{l}
\left(0.7-0.2 q^{1 / 3}\right) q^{1 / 3}, \quad q \leq 1, \\
1-\left(0.7-0.2 q^{-1 / 3}\right) q^{-1 / 3}, \quad q>1.
\end{array}\right.
\end{equation}

\subsection{The wind-accretion process}

In the present work, we employ the Reimers wind before the donor evolves to the AGB phase, and adopt the Blocker wind after the donor evolves to the AGB phase (\citealt{1975MSRSL...8..369R,1995A&A...299..755B}).
For the mass-accretion efficiency of WDs, we consider both the Bondi-Hoyle mass-accretion efficiency and the wind Roche-lobe overflow (WRLOF) mass-accretion efficiency, and adopted the larger one in the calculations.

(1) The Bondi-Hoyle accretion efficiency (\citealt{1944MNRAS.104..273B,1988A&A...205..155B}) is written as:
\begin{equation}
{\beta}_{\mathrm{acc}, \mathrm{BH}}=\frac{1}{\sqrt{1-e^2}}\left(\frac{G M_{\mathrm{WD}}}{v_{\mathrm{w}}^2}\right)^2 \frac{\alpha_{\mathrm{acc}}}{2 a^2\left(1+\frac{v_{\mathrm{orb}}^2}{v_{\mathrm{w}}^2}\right)^{1.5}},
\end{equation}
where $e$ is the orbital eccentricity (we assumed that binary orbit is circular and 
$e = 0$ ), $\alpha_{\mathrm{acc}}$ is the accretion efficiency parameter that is generally set as 1.5 in $\text{MESA}$, $v_{\mathrm{orb}}$ is the orbital velocity, 
$v_{\mathrm{w}}$ is the wind velocity, we set $v_{\mathrm{w}}$ to $5\, \mathrm{km\, s^{-1}}$, which is similar to \cite{2011ApJ...735L..31C}.
\cite{2013A&A...552A..26A} suggested that stellar wind velocity of AGB star is in the range $5\mathrm{-}30\, \mathrm{km \, s^{-1}}$ when the binary period is around $10^4\, \rm d$. 
A more detailed relationship between the wind velocity and the escape velocity can be seen in \cite{2006MNRAS.367..186E}.

(2) WRLOF occurs when the wind acceleration radius of AGB star is larger than the Roche-lobe radius, during which WD can accrete material in the wind-accretion zone through the inner Lagrangian point (\citealt{2012BaltA..21...88M,2013A&A...552A..26A}). 
The WRLOF mass-accretion efficiency can be expressed as 
\begin{equation}
\beta_{\mathrm{acc},\mathrm{WRLOF}}=\frac{25}{9} q^{-2}\left[-0.284\left(\frac{R_{\mathrm{d}}}{R_{\mathrm{L}}}\right)^2+0.918 \frac{R_{\mathrm{d}}}{R_{\mathrm{L}}}-0.234\right].
\end{equation}
in which $q = M_{2} / M_{\mathrm{WD}}$, ${R_{\mathrm{d}}}=0.5 {R_*}({T_{\mathrm{d}}}/{T_{\mathrm{eff}}})^{-(4+p)/{2}}$ is the dust formation radius, ${R_*}$ is the donor radius, $T_{\mathrm{d}}$ is the condensation temperature of the dust, $T_{\mathrm{eff}}$ is the effective temperature of the donor, $p$ is a parameter characterizing the wavelength dependence of the dust opacity.
According to previous studies of \cite{2007ASPC..378..145H} and \cite{2013A&A...552A..26A}, we adopted $T_{\mathrm{d}} = 1500\, \mathrm{K} $ and $p = 1$ in this work.
It is worth noting that we artificially limit the mass-accretion efficiency of the WD (${\beta}_{\mathrm{acc} , \mathrm{BH}}\leq 1.0, \ {\beta}_{\mathrm{acc}, \mathrm{WRLOF}}\leq 0.8 $), because the WDs cannot completely accrete the matter ejected from the donors (e.g. \citealt{2009MNRAS.396.1086L,2019MNRAS.485.5468I,2021MNRAS.503.4061W}). 

\subsection{Mass-growth rate of WDs}

Generally, the WD mass-growth rate remains controversial, especially for the recurrent nova outbursts during the mass-transfer process (e.g. \citealt{2005ApJ...623..398Y,2007ApJ...663.1269N,2017ApJ...847...99M}). 
In this work, we use the prescription provided by \cite{1999ApJ...519..314H} to calculate the WD mass-growth rate, which can be written as
\begin{equation}
\dot{M}_{\mathrm{WD}}=\eta_{\mathrm{He}} \eta_{\mathrm{H}} \dot{M}_{\mathrm{acc}},
\end{equation}
in which $\eta_{\mathrm{H}}$ is the mass-accumulation efficiency for H-shell burning (e.g. \citealt{2010MNRAS.401.2729W}), $\eta_{\mathrm{He}}$ is the mass-accumulation efficiency for He-shell flashes (\citealt{2004ApJ...613L.129K}). 

When the WD mass-accretion rate is larger than a critical mass-accretion rate ($\dot{M}_{\mathrm{cr}}$), we assume that the WD accumulates H-rich matter at the rate of $\dot{M}_{\mathrm{cr}}$, the rest of matter wound be blown away in the form of the optically thick wind (e.g. \citealt{1982ApJ...253..798N,1994ApJ...437..802K,1996ApJ...470L..97H}).
The critical mass-accretion rate is 
\begin{equation}
\dot{M}_{\mathrm{cr,H}}=5.3 \times 10^{-7} \frac{1.7-X}{X}\left(M_{\mathrm{WD}}-0.4\right) \mathrm{M}_{\odot} \mathrm{yr}^{-1},
\end{equation}
in which $X$ is the H mass fraction, and $M_{\mathrm{WD}}$ is the mass of WDs in units of $M_{\odot}$. 
The mass-accumulation efficiency of hydrogen can be expressed as follows:
\begin{equation}
\eta_{\mathrm{H}}=\left\{\begin{aligned}
\frac{\dot{M}_{\mathrm{cr}, \mathrm{H}}}{\dot{M}_{\mathrm{acc}}},\ & \dot{M}_{\mathrm{acc}}>\dot{M}_{\mathrm{cr}, \mathrm{H}},\\
1,\ & \frac{1}{2} \dot{M}_{\mathrm{cr}, \mathrm{H}} \leq \dot{M}_{\mathrm{acc}} \leq \dot{M}_{\mathrm{cr}, \mathrm{H}} ,\\
1,\ & \frac{1}{8} \dot{M}_{\mathrm{cr}, \mathrm{H}} \leq \dot{M}_{\mathrm{acc}} \leq \frac{1}{2} \dot{M}_{\mathrm{cr}, \mathrm{H}} ,\\
0,\  & \dot{M}_{\mathrm{acc}} < \frac{1}{8} \dot{M}_{\mathrm{cr}, \mathrm{H}} .
\end{aligned}\right.
\end{equation}
where $\eta_{\mathrm{H}}$ can be divided into three cases: 
(1) The optically thick wind phase. The accreted hydrogen stably burns into He at the rate of $\dot{M}_{cr}$ when $\dot{M}_{\mathrm{acc}}>\dot{M}_{\mathrm{cr}, \mathrm{H}}$. 
(2) The stable H-shell burning phase. When ${1}/{2} \dot{M}_{\mathrm{cr}, \mathrm{H}} \leq \dot{M}_{\mathrm{acc}} \leq \dot{M}_{\mathrm{cr}, \mathrm{H}}$, the WD can accumulate all accreted material. 
(3) The weak H-shell flash phase. The weak H-shell flash occurs and no matter is lost from the binaries when ${1}/{8} \dot{M}_{\mathrm{cr}, \mathrm{H}} \leq \dot{M}_{\mathrm{acc}} \leq {1}/{2} \dot{M}_{\mathrm{cr}, \mathrm{H}}$.
(4) The strong H-shell flash phase. When $\dot{M}_{\mathrm{acc}} < {1}/{8} \dot{M}_{\mathrm{cr}, \mathrm{H}}$, strong H-shell flash 
causes WD not to accumulate any material to increase its mass.

\section{RESULTS} \label{3. RESULTS}

\begin{figure}
	\begin{center}
		\centering
		\epsfig{file=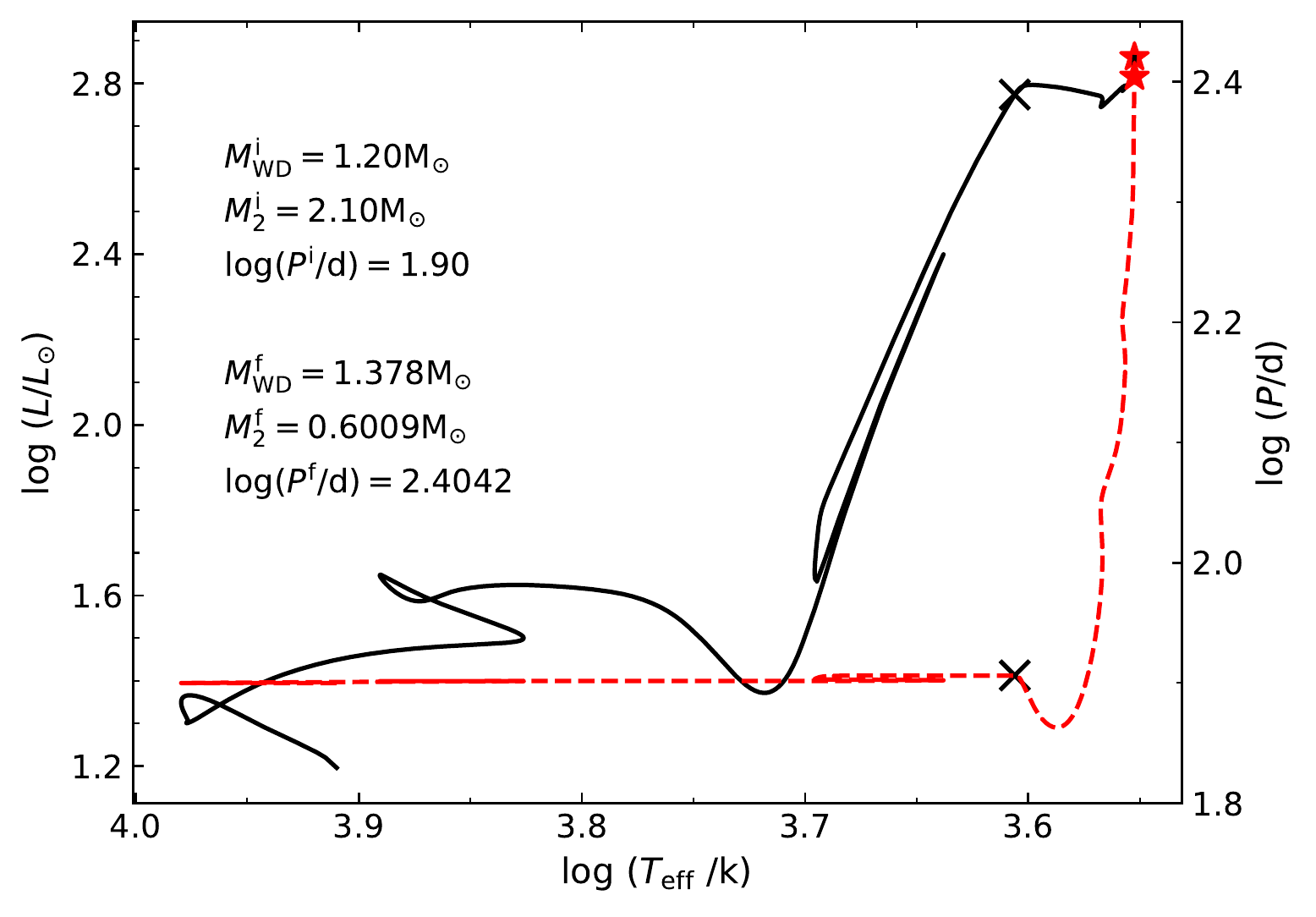,angle=0,width=7cm }
		\centering
		\epsfig{file=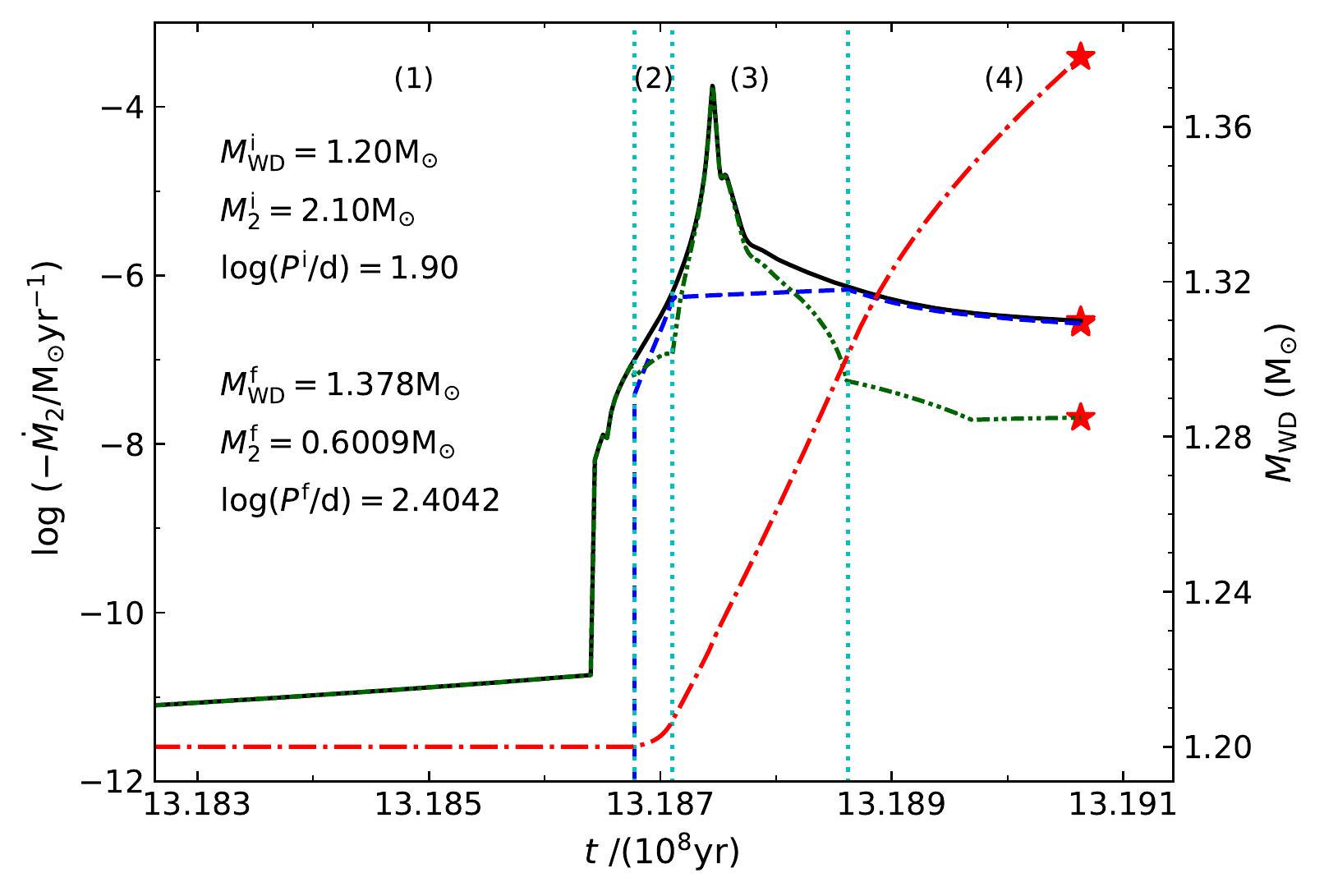,angle=0,width=7.4cm }	
		\caption{A typical binary evolution for producing an SN \romeI a through RLOF. In the left figure, the black solid curve stands for the evolutionary track of the mass donor in the H–R diagram, and the red dash-dotted curve shows the evolution of the orbital periods. The black crosses stand for the start of mass-transfer. In the right figure, the evolution of WD mass-accretion rate ($\dot{M}_\mathrm{{acc}}$), WD mass-growth rate ($\dot{M}_{\mathrm{WD}}$), binary mass-loss rate ($\dot{M}_\mathrm{loss}$) and WD mass ($M_{\mathrm{WD}}$) as a function of time are shown as black solid, blue dashed, green dash-dotted and red dash-dotted curves, respectively. The asterisks stand for the position where an SN \romeI a explosion occurs.}
	\end{center}
\end{figure}

\begin{figure}
	\begin{center}
		\centering
		\epsfig{file=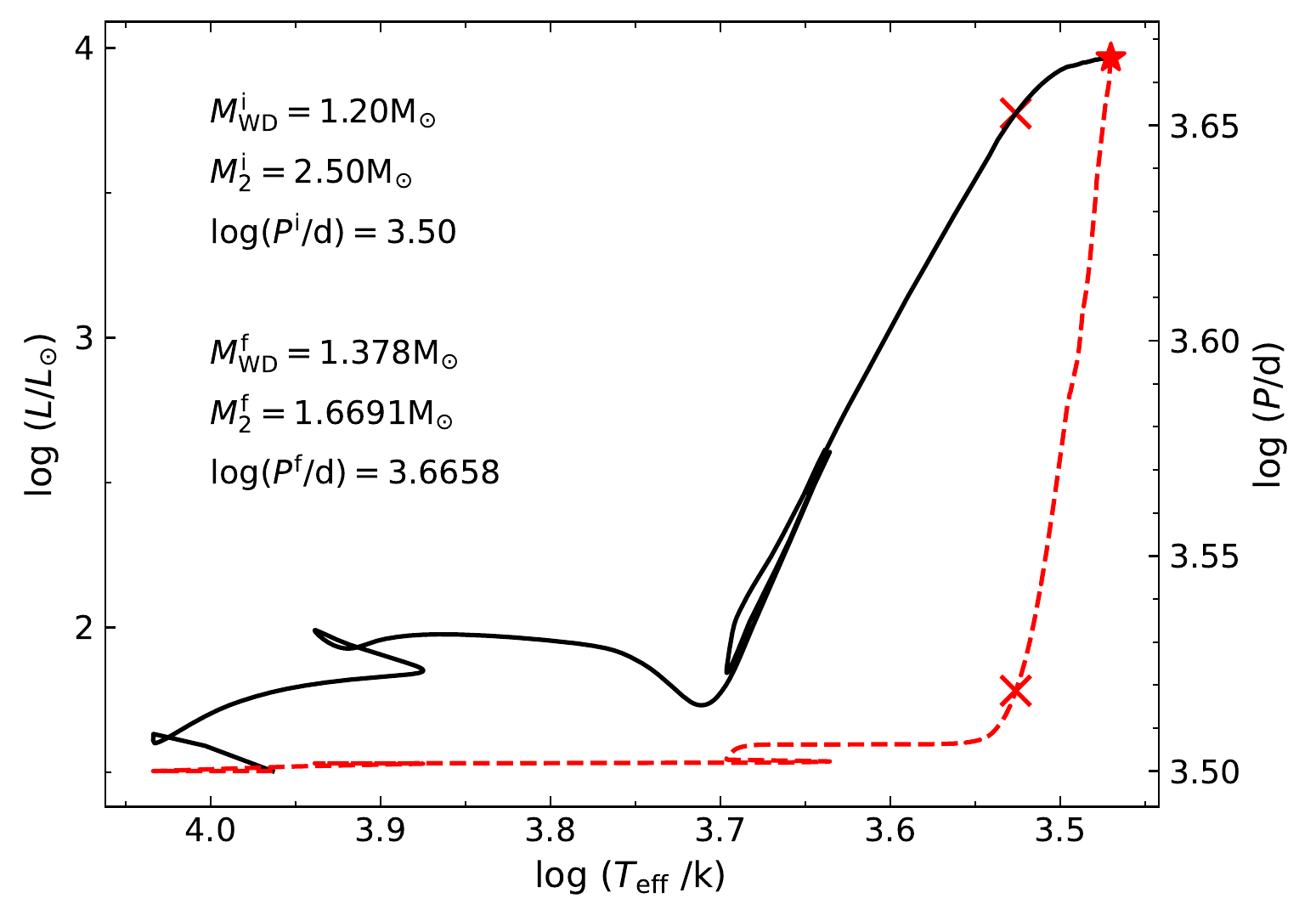,angle=0,width=7cm }
		\centering
		\epsfig{file=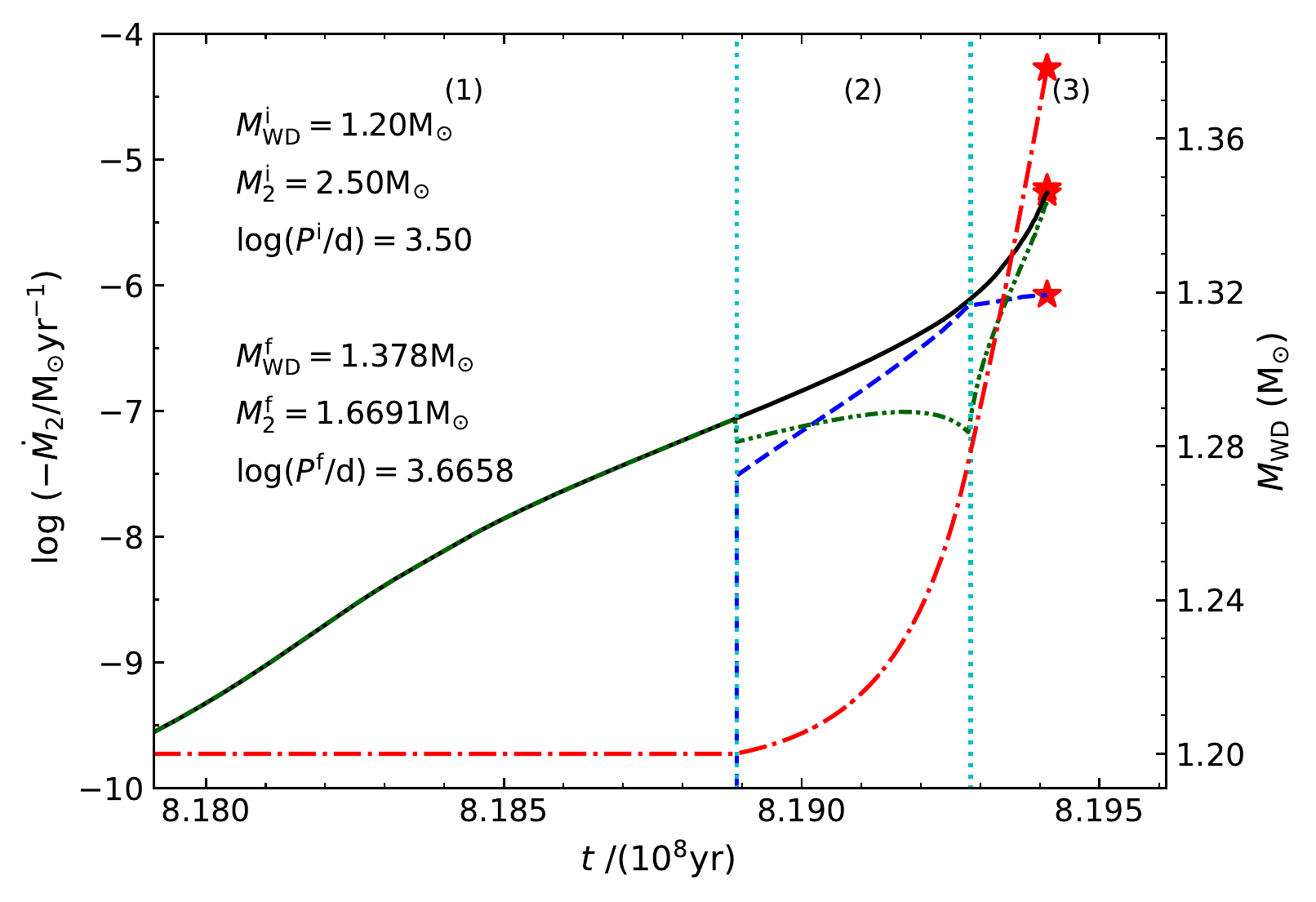,angle=0,width=7.4cm }	
		\caption{Similar to Fig.\,1, but for a typical binary evolution for producing an SN \romeI a through stellar wind-accretion. The red crosses in the left figure stand for the beginning of WD mass growth.In the right figure, the evolution of WD mass-accretion rate ($\dot{M}_\mathrm{{acc}}$), WD mass-growth rate ($\dot{M}_{\mathrm{WD}}$), binary mass-loss rate ($\dot{M}_\mathrm{loss}$) and WD mass ($M_{\mathrm{WD}}$) as a function of time are shown as black solid, blue dashed, green dash-dotted and red dash-dotted curves, respectively. 
            The WD mass-accretion rate in this case is equal to the donor mass-loss rate multiplied by mass-accretion efficiency.
            The asterisks stand for the position where an SN \romeI a explosion occurs.}
	\end{center}
\end{figure}

In order to explore the parameter space for producing SNe \romeI a, we evolved about 600 WD$+$AGB systems, for which the initial masses of the WDs are in the range from 1.15 to $1.25\, \rm {M_{\odot}}$, the initial masses of the donors are in the range of $1.8\mathrm{-}3.0\, \rm {M_{\odot}}$. The initial orbital periods are in the range of $25\mathrm{-}25000\, \mathrm{d}$;
the donor in a binary with a shorter period will fill its Roche-lobe in the RG phase, and the binary with longer period will experience mass-transfer with a high rate and lose so much mass via the optically thick wind that the WD cannot increase its mass to $M_{\mathrm{Ch}}$.

\subsection{Examples of binary evolution calculations}

Fig.$\,$1 presents a typical example of binary evolution computations with the initial parameters of $(M_{\rm WD}^{\rm i}, M_2^{\rm i}, \mathrm{log} (P^{\mathrm{i}}/\mathrm{d})) = (1.2, 2.1, 1.9)$, in which the AGB donor could fill its Roche-lobe.
The mass-transfer process can be divided into four parts before the WD increases its mass to $M_{\mathrm{Ch}}$: 
(1) The strong H-shell flash phase.
The AGB donor fills its Roche-lobe at about $t = 1.3186\, \mathrm{Gyr}$. 
In this case, $\dot{M}_{\mathrm{acc}}$ is less than $\frac{1}{8} \dot{M}_{\mathrm{cr}}$ and no matter can be accumulated onto the WD because of the strong H-shell flash.
(2) The stable H-shell burning or the weak H-shell flash phase.
After about $3.78\times 10^{4}\, \mathrm{yr}$, $\dot{M}_{\mathrm{acc}}$ increases to greater than $\frac{1}{8} \dot{M}_{\mathrm{cr}}$ and the WD begins to increase its mass.
(3) The optically thick wind phase.
After about $3.26\times 10^{4}\, \mathrm{yr}$, $\dot{M}_{\mathrm{acc}}$ continues to increase to be larger than $\dot{M}_{\mathrm{cr}}$. 
In this case, we assume that the H-rich matter burns into He at a rate of $\dot{M}_{\mathrm{cr}}$ and the rest of matter is assumed to be blown away via the optically thick wind on the surface of WDs.
The mass-accretion rate has a peak around  $t = 1.3187\times 10^9\, \mathrm{yr}$, which is corresponding to the moment when the mass ratio $q = 1$. 
After that, the donor turns to be less massive than the accretor and the mass-transfer rate of RLOF starts to decrease.
(4) The stable H-shell flash phase.
After about $1.52\times 10^5\, \mathrm{yr}$, $\dot{M}_{\mathrm{acc}}$ decreases to be lower than $\dot{M}_{\mathrm{cr}}$. This case is similar to Phase\,(2).
At about $t = 1.3190\, \mathrm{Gyr}$, the WD increases its mass to $M_{\mathrm{Ch}}$ and an SN \romeI a is assumed to occur.
At this moment, the donor evolves to a $0.6009\, \mathrm{M}_{\odot}$ AGB star and the orbital period is $0.69\,\mathrm{yr}$. 

Fig.$\,$2 shows a typical example of binary evolution computations with the initial parameters of $(M_{\rm WD}^{\rm i}, M_2^{\rm i}, \mathrm{log} (P^{\mathrm{i}}/\mathrm{d})) = (1.2, 2.5, 3.5)$, in which the WD increase its mass through the wind-accretion process. 
The mass-transfer process can be also divided into three parts before the WD increases its mass to $M_{\mathrm{Ch}}$: 
(1) The strong H-shell flash phase. 
Before about $t = 8.1889\times 10^8\, \mathrm{yr}$, $\dot{M}_{\mathrm{acc}}$ is less than $\frac{1}{8} \dot{M}_{\mathrm{cr}}$.
The AGB stellar wind works but the WD do not increase mass because of the strong H-shell flash.
(2) The stable H-shell burning or the weak H-shell flash phase. 
At about $t = 8.1889\times 10^8\, \mathrm{yr}$, $\dot{M}_{\mathrm{acc}}$ is larger than $\frac{1}{8} \dot{M}_{\mathrm{cr}}$ and WD begins to increase its mass from stellar wind of AGB donor.
(3) The optically thick wind phase.
After about $3.92\times 10^5\, \mathrm{yr}$, $\dot{M}_{\mathrm{acc}}$ increases to be larger than $\dot{M}_{\mathrm{cr}}$ and WD accumulates matter as a rate of $\dot{M}_{\mathrm{cr}}$.
At about $t = 8.1941\times10^8 \, \mathrm{yr}$, the WD increases its mass to $M_{\mathrm{Ch}}$ and an SN \romeI a is supposed to occur.
At this moment, the donor evolves to a $1.6691\, \mathrm{M}_{\odot}$ AGB star and the orbital period is $12.69\,\mathrm{yr}$.

\begin{figure}
	\begin{center}
		\centering
		\epsfig{file=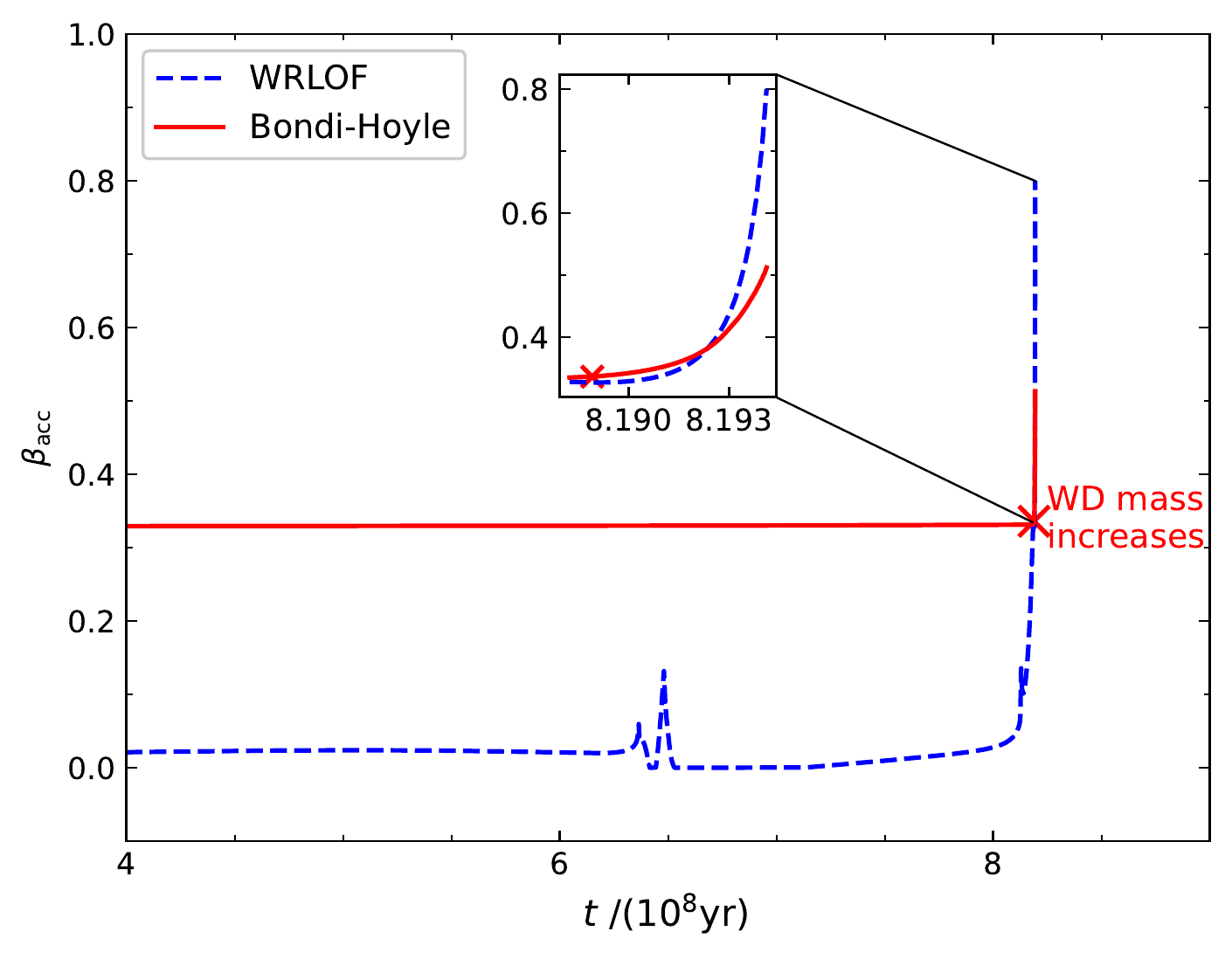,angle=0,width=10cm}
		\caption{The comparison of Bondi-Hoyle accretion efficiency and WRLOF accretion accretion for the wind-accretion case shown in Fig.\,2. Red solid and green dashed curves stand for Bondi-Hoyle accretion efficiency and WRLOF accretion accretion, respectively. The red cross stands for the beginning of WD mass growth.}
	\end{center}
\end{figure}

Fig.$\,$3 shows the comparison of the Bondi-Hoyle accretion efficiency and the WRLOF accretion efficiency for the wind-accretion case shown in Fig.$\,$2.
From this figure, we can see that the Bondi-Hoyle accretion efficiency works before the donor evolves to AGB phase.
Note that the curve of WRLOF accretion efficiency has two peaks around $t = 6.5\times 10^8\, \mathrm{yr}$, which corresponding to the Hertzsprung-Gap phase and the RGB phase.
The donor expands rapidly during these two phases, leading to the decrease of its effective temperature. 
In this case, the value of the $R_{\mathrm{d}}$ decreases and ${\beta}_{\mathrm{acc}, \mathrm{WRLOF}}$ significantly increases (see Eq.\,7).
When the donor evolves to the AGB phase at $t = 8.0\times 10^8\, \mathrm{yr}$, it expands quickly and ${\beta}_{\mathrm{acc}, \mathrm{WRLOF}}$ significantly increases over ${\beta}_{\mathrm{acc}, \mathrm{BH}}$, during which the WRLOF accretion efficiency starts to work.

\begin{figure}
	\begin{center}
		\centering
		\epsfig{file=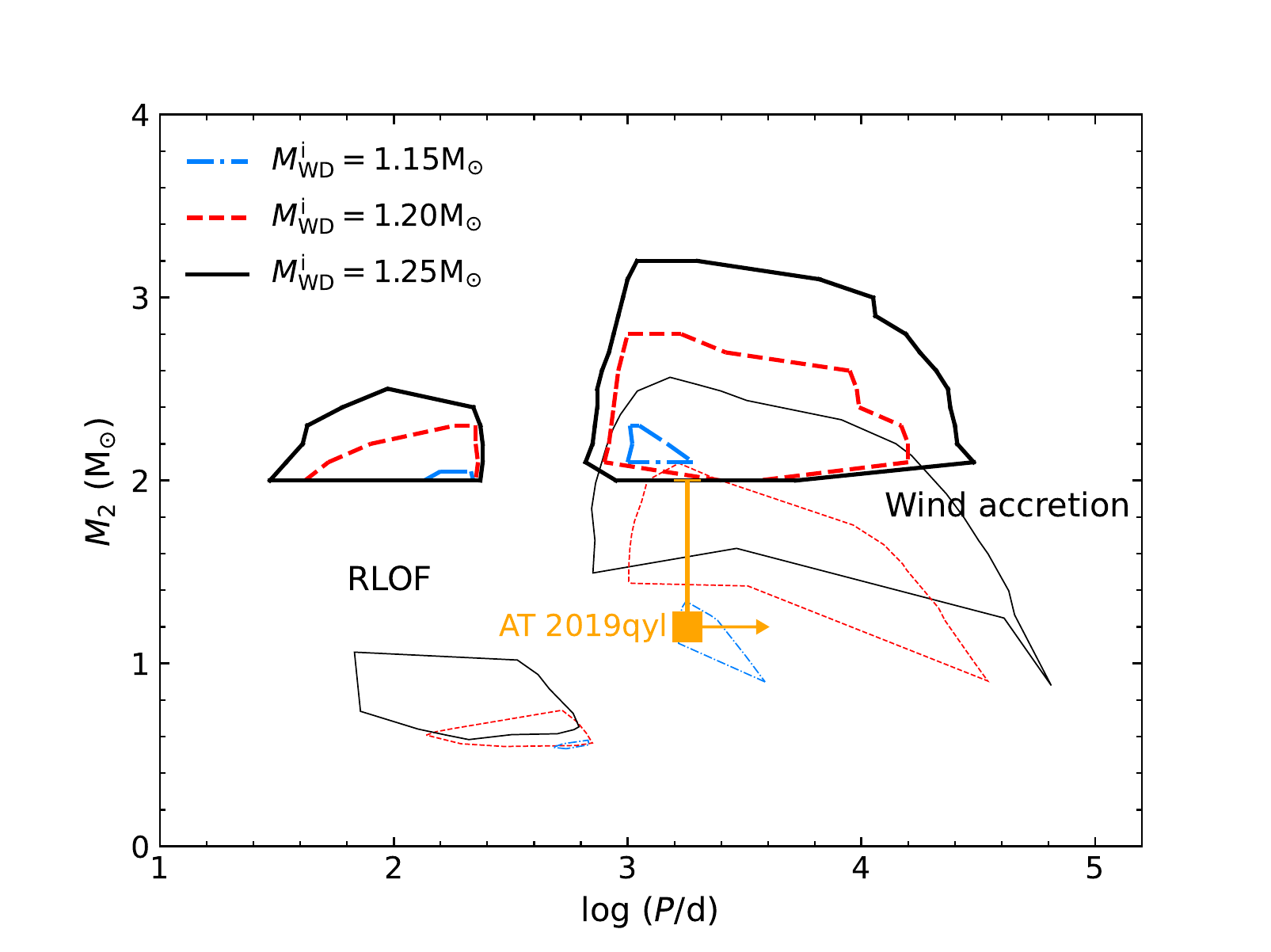,angle=0,width=13cm}
		\caption{ Initial and final regions of WD$+$AGB systems in their orbital period$-$donor mass ($\mathrm{log}\ P-M_{2}$) plane for producing SNe \romeI a with different intital WD masses. The thick and thin lines represent initial and final parameter space, respectively. The left and right contours represent the RLOF cases and wind-accretion cases, respectively. The data for AT$\,$2019qyl are taken from \cite{2021ApJ...920..127J}. 
		}
	\end{center}
\end{figure}

\subsection{Parameter space for producing SNe Ia}

Fig.$\,4$ shows the initial and final contours of WD$+$AGB systems for producing SNe \romeI a with the initial WD masses of 1.15, 1.20 and 1.25$\,\rm M_{\odot}$.
The initial donor masses for producing SNe \romeI a are larger than 2$\,\rm M_{\odot}$.
The intermediate-mass stars will develop convective envelopes when their masses decrease to be less than $1.5\,\rm M_{\odot}$, after which the magnetic braking should work (e.g. \citealt{1983ApJ...275..713R,2015ApJS..220...15P,2020ApJ...900L...8C,2021ApJ...909..174D,2022MNRAS.515.2725G}).
In the present work, we ignore the magnetic braking, even when the donors evolve into low-mass stars with masses less than $1.5\,\rm M_{\odot}$. 
From this figure, we can see that as the initial WD mass increases, the initial parameter spaces of the RLOF case expands to the upper left, and the wind-accretion case expands upper right.
It is notable that the position of AT$\,$2019qyl can be basically covered by the final contours of the wind-accretion case, which indicates that AT$\,$2019qyl is a strong progenitor candidate of SNe \romeI a.

The surrounding boundaries of initial parameter space are determined by different reasons. 
The binaries beyond the upper boundaries cannot produce SNe \romeI a because too much material is lost via optically thick wind during the mass-transfer phase due to the large mass ratios.
The lower boundaries of the two contours are set by the less massive donors and the low mass-transfer rate, in which the WDs cannot increase their masses to $M_{\mathrm{Ch}}$. 
The donors in binaries beyond the left RLOF boundaries will fill their Roche-lobes at the RGB phase.
The binaries beyond the right wind-accretion boundaries and between the two contours are caused by the fact that these binaries have experienced relatively fast mass-transfer processes with and thus lost too much mass through the optically thick wind.

\begin{figure}
	\begin{center}
		\centering
		\epsfig{file=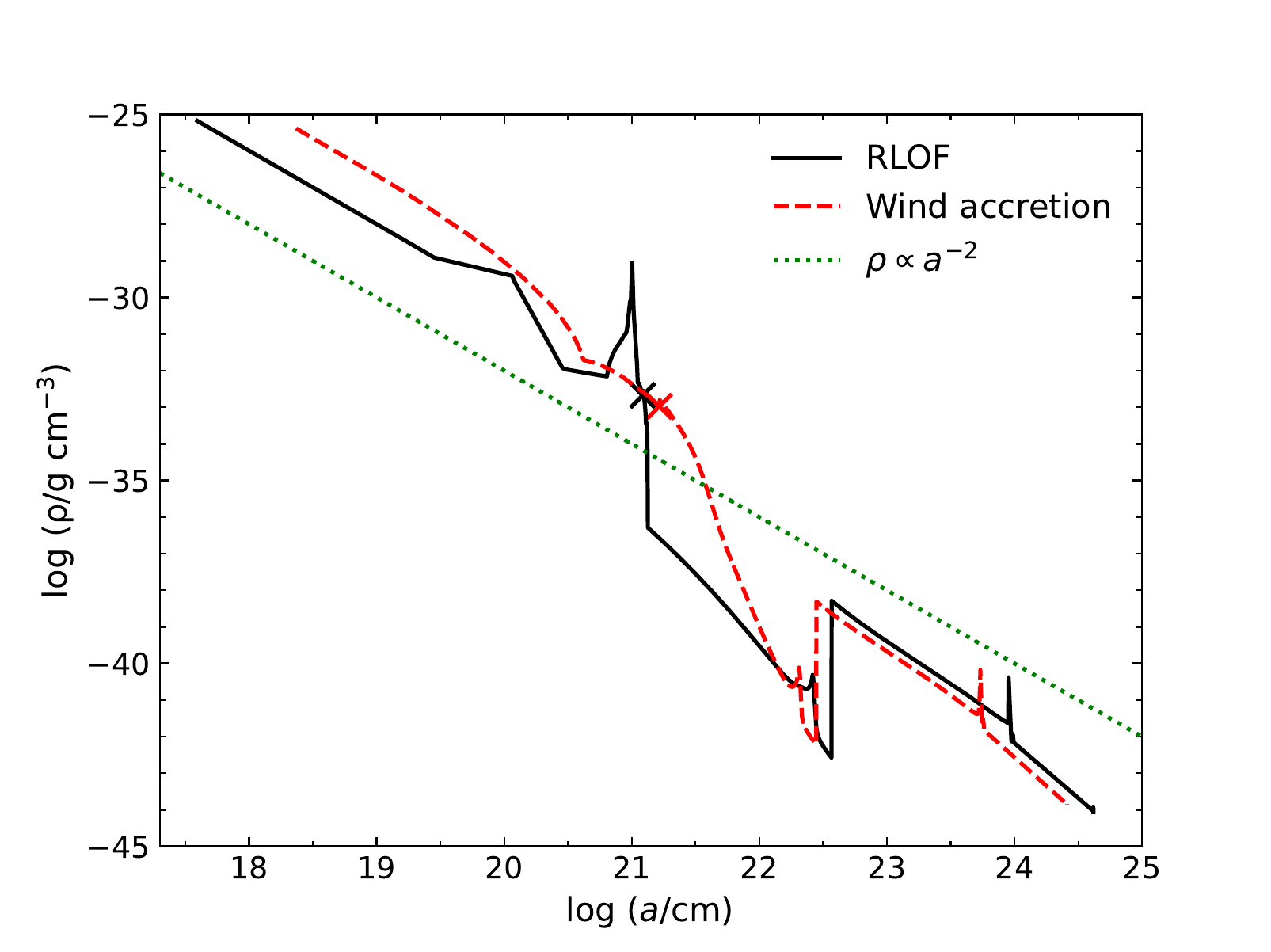,angle=0,width=10.5cm }
		\caption{ The density distribution comparison of CSM at the moment of SN \romeI a explosion for the RLOF case shown in Fig.$\,1$ and wind-accretion case shown in Fig.$\,2$. 
                    The black and red crosses stand for the start of mass-transfer and the WD mass growth, respectively.}
		
	\end{center}
\end{figure}

\subsection{Density distribution of CSM}

Similar to \cite{2019MNRAS.488.3949M}, the assumptions for the wind velocity of the lost material during the mass-transfer process are shown as follows:
(1) In the stable H-shell burning phase, we assume that about 1\% of transferred mass escaping from the outer Lagrangian point and the wind velocity is supposed to approximately equal to the orbital velocity ($\sim$100 $\mathrm{km\, s^{-1}}$ ) (\citealt{1996ChA&A..20..175H,1999A&A...343..455D}).
This assumption is only used to estimate the density distribution of CSM. In the binary evolution calculations, we do not consider this mass escape from the outer Lagrangian point.
(2) In the weak H-shell flash phase, the wind velocity is assumed to be similar to that of novae, which is assumed to be about 1000 $\mathrm{km\, s^{-1}}$. 
(3) In the optically thick wind phase, the wind velocity is assumed as the escape velocity at the radius of H-envelope and approximated as a speed of approximately 1000 $\mathrm{km\, s^{-1}}$.
After these simplifications, the density of CSM can be expressed as 
$\rho = {\dot{M}_\mathrm{loss}}/ ({4\pi a^{2} V_\mathrm{{loss}}})$, where $\dot{M}_\mathrm{loss}$ is the mass-loss rate of the binary, $a$ is the distance from the binary, and $V_\mathrm{{loss}}$ is the wind velocity of the lost material. 

Fig.$\,$5 presents the density distribution of CSM for the evolutionary cases in Fig.\,1 (RLOF) and Fig.\,2 (wind accretion) when the WD masses increase to $M_{\mathrm{Ch}}$.
From this figure, we can see that the distribution basically meets $\rho \propto a^{-2}$.
Note that the CSM in the region of $\mathrm{log}\, a \gtrsim 22 $ is similar for these two cases, because neither of their donors fill their Roche-lobes and the mass-loss originates from the stellar wind of the donors.
There is a small peak at $\mathrm{log}\, a \approx 23.8$.
At this time, the two donors evolve to their RGB phase and the stellar wind becomes stronger. 
They evolve to AGB phase when $\mathrm{log}\, a \approx 22.5$. 
We can also see that there is a peak in the curve of RLOF case around $\mathrm{log}\, a \approx 21$. 
The donor fills its Roche-lobe at this time and the mass-transfer rate increases rapidly. 
In this case, a large amount of matter lost from the binary in the form of the optically thick wind.

\section{DISCUSSION AND SUMMARY} \label{4. DISCUSSION AND SUMMARY}

The CSM forms during the mass-transfer process will interact with SN ejecta, which would generate radio synchrotron emission and X-ray emission.
The physical processes and characteristic features of the interactions have been well studied (e.g. \citealt{1998ApJ...499..810C,2006ApJ...651..381C,2012ApJ...758...81M}).
\cite{2016A&A...588A..88M} found that the X-ray and radio flux may be high enough to be detected for a nearby SN \romeI a from a WD$+$AGB system.
From Fig.$\,4$, we can infer that the masses of CSM at the moment of SNe \romeI a explosion in RLOF cases and wind-accretion cases are in the range of $0.85\mathrm{-}1.69\,\mathrm{M_{\odot}} $ and $0.15\mathrm{-} 1.29\,\mathrm{M_{\odot}}$, respectively.
According to binary evolution calculations, we can summarize that the mass-loss rate at the moment of SNe \romeI a explosion is in the range of $8.38\times10^{-9}\mathrm{-} 3.61\times10^{-6}\, \mathrm{M_{\odot}yr^{-1}}$ for the RLOF cases, and $2.64\times10^{-8} \mathrm{-} 5.05\times10^{-5}\, \mathrm{M_{\odot}yr^{-1}}$ for the wind-accretion cases.

Unlike previous studies, we found that the semidetached WD$+$AGB binaries can also produce SNe \romeI a in the present work.
In the RLOF process, the integrated mass-transfer prescription is more physical and suitable for semidetached binaries with giant donors.
This prescription is based on laminar mass overflow and the stellar state equation that obeys the adiabatic power law.
When the donor fills its Roche-lobe, the mass-transfer rate is lower than that of previous models, resulting in that the WDs can accumulate more material through stable RLOF process (for more discussions see \citealt{2019A&A...622A..35L}).

It is worth noting that the CD model for producing SNe \romeI a also involves the WD$+$AGB systems.
In the CD model, the merger of a WD with the hot CO-core of an AGB star during or after a common envelop phase would produce an SN \romeI a (e.g. \citealt{2011MNRAS.417.1466K,2012MNRAS.419.1695I,2013MNRAS.428..579I,2014MNRAS.437L..66S,2015MNRAS.450.2948A}).
\cite{2015MNRAS.447.2568T} estimated that at least 20 percent of all SNe \romeI a are produced by this channel. 
Recently, \cite{2022arXiv221104353S} suggested that the merger of a WD with the hot CO-core of an He subgiant can explain the He-rich CSM of SN 2020eyj under the CD model.
In this model, the common envelope ejection will form one or multiple shells. 
\cite{2013MNRAS.431.1541S} argue that the multiple shells of CSM in SN \romeI a PTF 11kx can be explained by a merger of WD and the hot core of an AGB star.
But in this work, the CSM has a continuous distribution that basically meets $\rho \propto a^{-2}$, which is the basic difference for the CSM distribution between the SD model and the CD model.
It has been suggested that the Kepler’s SNR may be the result of SN \romeI a explosion in SD model with an AGB donor (\citealt{2012A&A...537A.139C}).

It has been suggested that an accretion disk is possibly formed around the WD during the mass-transfer process, and the accretion disk may become thermally unstable when the effective temperature in the disk falls below the hydrogen ionization temperature (e.g. \citealt{1996ApJ...464L.139V,1997ApJ...484..844K,2001NewAR..45..449L}).
Some previous studies have investigated the influence of the thermally unstable accretion of WD binaries (e.g. \citealt{2009A&A...495..243X,2010MNRAS.401.2729W,2010RAA....10..235W}).
After considering the disk instability, it has been found that the mass-accumulation efficiency of WD can be significantly improved and the systems with less-massive donors can also produce possible SNe \romeI a, which would be helpful to explain the SNe \romeI a with long delay times (\citealt{2007ApJ...658L..51C,2009A&A...495..243X,2010MNRAS.401.2729W,2010RAA....10..235W}).
In this case, we can infer that the lower boundaries of initial parameter space for producing SNe \romeI a would expand downwards because of the larger mass-accumulation efficiency of WD if the accretion-disk instability is considered in WD$+$AGB binaries.
\cite{2022ApJ...941L..33A} compared the evolution of non-magnetic and magnetized WD$+$RG binaries, and found that the accretion would occur on the two small polar caps of the WDs, which may potentially suppress nova outbursts. 
They suggested that the WD$+$RG binaries with shorter orbital periods and lower donor masses in the initial parameter space could produce SNe \romeI a if the magnetic confinement is considered.
Therefore, we can speculate that magnetic confinement would have a similar effect on the AGB donor channel for producing SNe \romeI a .

In the present work, the accretor is treated as a mass point, and thus the provided parameter space is also useful if the accretor is an oxygen-neon (ONe) WD, which may evolve to the accretion-induced collapse (AIC) events.
Unlike CO WDs, massive ONe WDs in close binaries may experience the AIC process when their masses approach to $M_\mathrm{{Ch}}$, which would lead to the formation of neutron star systems (e.g. \citealt{1986ApJ...305..235T,1987Natur.329..310M,1990ARA&A..28..183C}).
The neutron stars can be spun up after the donors refill their Roche-lobe, which is a possible path for the formation of millisecond pulsars (e.g. \citealt{1991PhR...203....1B,2013A&A...558A..39T,2012ApJ...756...85S}). 
In this case, the symbiotic systems may also evolve to NS systems via the AIC process.
\cite{2022MNRAS.510.6011W} investigated the formation of millisecond pulsars through the RG donor channel, and found that there exists an anticorrelation between the final neutron star mass and the final orbital period based on this channel.
\cite{2023MNRAS.519.1327A} investigated the evolution of non-magnetic or magnetized ONe WDs$+$RG binaries, and found the initial parameter space shifts to be lower and narrower after considering the influence of the magnetic field.

In this work, we studied the formation of SNe \romeI a from the semidetached and detached WD$+$AGB systems.
We found that the semidetached WD$+$AGB system is a possible path for the formation of SNe \romeI a after a more physical mass-transfer method is adopted.
In addition, we provided the parameter space of the semidetached and detached WD$+$AGB systems for the formation of SNe Ia.
We also compared the density distribution of CSM from these two cases.
We suggest that AT$\,$2019qyl is a strong candidate for the progenitors of SNe \romeI a with AGB donors.
In order to understand the AGB donor channel for the formation of SNe \romeI a, further numerical research on the mass-transfer prescription for semidetached binaries with giant donors are needed, and large samples of observed WD+AGB systems are expected.

\begin{acknowledgements}
We acknowledge the useful comments and suggestions from the referee.
This study is supported by the National Natural Science Foundation of China (Nos 12225304, 12273105 and 11903075), the National Key R\&D Program of China (Nos 2021YFA1600404, 2021YFA1600403 and 2021YFA1600400), the Western Light Project of CAS (No. XBZG-ZDSYS-202117), the science research grants from the China Manned Space Project (No. CMS-CSST-2021-A12), the Youth Innovation Promotion Association CAS (No. 2021058), the Yunnan Fundamental Research Projects (Nos 202001AS070029, 202001AU070054, 202101AT070027, 202101AW070047 and 202201BC070003), the Frontier Scientific Research Program of Deep Space Exploration Laboratory (No. 2022-QYKYJH-ZYTS-016) and International Centre of Supernovae, Yunnan Key Laboratory (No. 202302AN360001).
\end{acknowledgements}

\bibliography{1bib}
\bibliographystyle{raa}

\end{document}